\documentclass{article}
\pdfoutput=1 

\usepackage{amsmath,amssymb}

\usepackage{boldline}
\usepackage{microtype}
\usepackage{graphicx}
\usepackage{subcaption}
\usepackage{booktabs} 
\usepackage{stmaryrd}
\usepackage{enumitem}
\usepackage{caption}
\usepackage[symbol]{footmisc}

\newtheorem{defn}{Definition}

\newtheorem{thm}{Theorem}

\usepackage{hyperref}

\newcommand{\ip}[2]{\langle #1,#2 \rangle}

\DeclareMathOperator*{\argmin}{argmin}

\newcommand{\tsim}[0]{\mathbb{S}^*}
\newcommand{\temb}[0]{\mathbf{E}^*}
\newcommand{\tkern}[2]{\phi(#1,#2)}
\newcommand{\emb}[0]{\mathbf{E}}
\newcommand{\psim}[0]{\mathbb{S}}
\newcommand{\alloc}[0]{\mathcal{A}}
\newcommand{\mem}[0]{\mathcal{M}}
\newcommand{\fcsm}[0]{f_{\mathcal{A}}}
\newcommand{\real}[0]{\mathbb{R}}

\usepackage{arxiv}

\begin{document}
\twocolumn[
\title{Semantically Constrained Memory Allocation (SCMA) for Embedding in Efficient Recommendation Systems}
\author{Aditya Desai\footnotemark[1] \\
  Department of Computer Science\\
  Rice University\\
  Houston, Texas \\
  \texttt{apd10@rice.edu} \\
   \And
Yanzhou Pan\footnotemark[1] \\
  Department of Computer Science\\
  Rice University\\
  Houston, Texas \\
  \texttt{yp24@rice.edu} \\
  \And
Kuangyuan Sun\\
  Department of Computer Science\\
  Rice University\\
  Houston, Texas \\
  \texttt{ks94@rice.edu} \\
    \And
Li Chou\\
  Department of Computer Science\\
  Rice University\\
  Houston, Texas \\
  \texttt{lchou@rice.edu} \\
    \And
Anshumali Shrivastava\\
  Department of Computer Science\\
  Rice University\\
  Houston, Texas \\
  \texttt{anshumali@rice.edu} 
}
\maketitle
]\footnotetext[1]{First two authors have equal contribution}



\setlength{\abovedisplayskip}{2pt}
\setlength{\belowdisplayskip}{2pt}

\begin{abstract}

Deep learning-based models are utilized to achieve state-of-the-art performance for recommendation systems. A key challenge for these models is to work with millions of categorical classes or tokens. The standard approach is to learn end-to-end, dense latent representations or embeddings for each token. The resulting embeddings require large amounts of memory that blow up with the number of tokens. Training and inference with these models create storage, and memory bandwidth bottlenecks leading to significant computing and energy consumption when deployed in practice. To this end, we present the problem of \textit{Memory Allocation} under budget for embeddings and propose a novel formulation of memory shared embedding, where memory is shared in proportion to the overlap in semantic information. Our formulation admits a practical and efficient randomized solution with Locality sensitive hashing based Memory Allocation (LMA). We demonstrate a significant reduction in the memory footprint while maintaining performance. In particular, our LMA embeddings achieve the same performance compared to standard embeddings with a 16$\times$ reduction in memory footprint. Moreover, LMA achieves an average improvement of over 0.003 AUC across different memory regimes than standard DLRM models on Criteo and Avazu datasets

\end{abstract}

\section{Introduction}
\vspace{-0.2cm}

Recommendation systems are at the core of business for companies such as Amazon, Facebook, NetFlix, and Google. These companies offer a wide array of products, movies, ads, etc. for customers to view and to choose from. Therefore, developing automated, intelligent, and personalized recommendation systems help guide customers to make more informed choices. It is worthwhile to note the significant monetary impact of recommendation model accuracy on the aforementioned companies. With the size of notional business, even an increase of 0.001  in accuracy/AUC metrics implies considerable gains in revenue and customer experience. However, categorical features (e.g., product history, pages liked, etc.) dominate most of the recommendation data \cite{DLRM19, cheng2016wide}, thereby posing new modeling challenges for learning. Following the lines of natural language processing \cite{word2vec,transformer-17}, state-of-the-art methods in recommendation models \cite{DLRM19, DCN17} have found success with mapping each of the category values in the feature to a dense representation. These representations are learned and stored in memory tables called {\em embedding} tables.

\textbf{Heavy embedding tables and memory bandwidth bottleneck}: 
Embedding tables store learned dense representation of each category value.  If the set of all categories is $S$ and the embedding dimension is $d$. The embedding table size would be $|S| {\times}d$. With the number of categories as large as tens of millions for each feature, embedding tables can take up over 99.9\% of the total memory. Namely, memory footprint can be multiple gigabytes or even terabytes~\cite{MDTrick19,tbsize1,tbsize2}. In practice, deploying these large models often requires the model to be decomposed and distributed across different machines due to memory capacity restrictions~\cite{archimpl}. Extensive memory utilization creates memory bandwidth issues due to the highly irregular locality of accesses making training and inference considerably slower~\cite{DLRM19}. This issue exacerbates further when multiple models need to be co-located on a single machine~\cite{archimpl}.

\textbf{Impact of improving memory usage of Embedding Tables}:
Memory consumption of embedding tables is a severe problem in recommendation models. Improving memory utilization can improve recommendation systems on various fronts. 1) It has been observed that larger embedding size in the model leads to better performance \cite{zhu2020fuxictr}. Better memory utilization would imply scope to train and deploy complex models. 2) Lower memory footprint will improve the training and inference speed. With changing consumer interests, recommendation data inherently suffers from concept shift \cite{conceptdrift}, requiring a frequent refresh of models. With faster training, models can be retrained more frequently, improving their average performance. Hence, memory utilization forms a critical problem requiring attention.

Deep learning recommendation model (DLRM)~\cite{DLRM19} gave rise to an increased interest in constructing more memory-efficient embeddings. Recent SOTA works include compositional embedding using complementary partition~\cite{QuoRemTrick19} and mixed dimension embeddings~\cite{MDTrick19}.
A simple memory sharing scheme for weight matrices in deep learning models was proposed by~\cite{hashtrick}. However, embeddings for tokens are weight matrices that have a structure for which we can reduce the memory burden in an intelligent manner by sharing memory for similar concept tokens. For example, if two tokens represent the concepts ``Nike'' and ``Adidas,'' we would expect the pair to share more weights compared to ``Nike'' and ``Jaguar.''
In this paper, we approach the problem of learning embedding tables under memory budget by solving a generic problem, which we refer to as {\em semantically constrained memory allocation} (SCMA) for embeddings.
SCMA is a hard combinatorial allocation problem since it involves millions of variables and constraints. Surprisingly, as we show later, that there is a very neat dynamic allocation of memory using locality sensitive hashing which solves SCMA in approximation. The memory allocation can be done in an online consistent manner for each token with negligible overhead.

This paper is organized as follows: we first formally describe the problem and notation in section 2, followed by a recap of hashing schemes in section 3 and a generic solution to the problem in section 4. Section 5 onward we focus on applying the solution to DLRM problem. Section 6 discusses some related work and we present experimental evaluation in section 7.
\vspace{-0.2cm}
\section{Semantically Constrained Memory Allocation (SCMA) for Embeddings} \label{sec:scma}

Let $S=\{0,1,\ldots,|S|{-}1\}$ denote the set of all values for all categorical features in the dataset. The embedding table $\emb \in R^{|S| \times d}$ is a matrix such that each row represents embeddings, each with dimension $d$, for all values in $S$. An embedding, $e_v$, for a particular value, is retrieved by $e_v=\emb[v,:]$ (see Figure \ref{fig:embedding-table}). The embedding table $\emb$ imposes a similarity structure $\psim$ w.r.t a similarity measure obtained by a kernel function $\tkern{.}{.}$. $\psim \in R^{|S|\times |S|}$ and each entry of this matrix $\psim[v_1, v_2] = \tkern{\emb[v_1,:]}{\emb[v_2,:]}$ also denoted as $\tkern{v_1}{v_2}$

The problem of end-to-end learning of full embedding table $\emb$ is well-known ~\cite{DLRM19, devlin2018bert, word2vec, goldberg2014word2vec, transformer-17}. In this paper, we consider the problem of learning $\emb$ under a memory budget $m$. Let $\mem$ be the memory such that $|\mem| = m$. If $|S|{\times}d > m$, then multiple elements of the embedding table \textit{have} to share memory locations in $\mem$. We formally define allocation function $\alloc$ as the mapping from elements of $\emb$ to the actual memory locations in $\mem$.

\begin{defn}[Allocation Function]
Allocation function for an embedding table $\emb \in R^{|S| \times d}$ using the memory $\mem$ with budget $|\mem|=m$, is defined as the following map.
\begin{align*}
    &\alloc : \{0,\ldots,|S|{-}1\} \rightarrow \{0,1\}^{d{\times}m} \quad\quad \textrm{s.t.}\\
    &\forall v \in S, \forall i \in \{0,\ldots,d{-}1 \}, \quad \sum_j(\mathcal{A}(v)[i,j]) = 1. 
\end{align*}
\vspace{-0.5cm}
\end{defn}
The $i^{th}$ row of the matrix output by $\alloc$, for any value $v$ is a one-hot encoding of the location to which the $i^{th}$ element of the embedding vector for $v$ maps to in $\mathcal{M}$. Using the allocation function, we can retrieve the embedding by,
\begin{align*}
    \emb[v,i] = \mem[\alloc(v)[i,:]]
\end{align*}
 where we assume mask-based retrieval on $\mathcal{M}$. We next define the notion of {\em shared memory} between two embeddings under a allocation function $\alloc$.
 
 \begin{figure}
    \centering
    \includegraphics[scale=0.5]{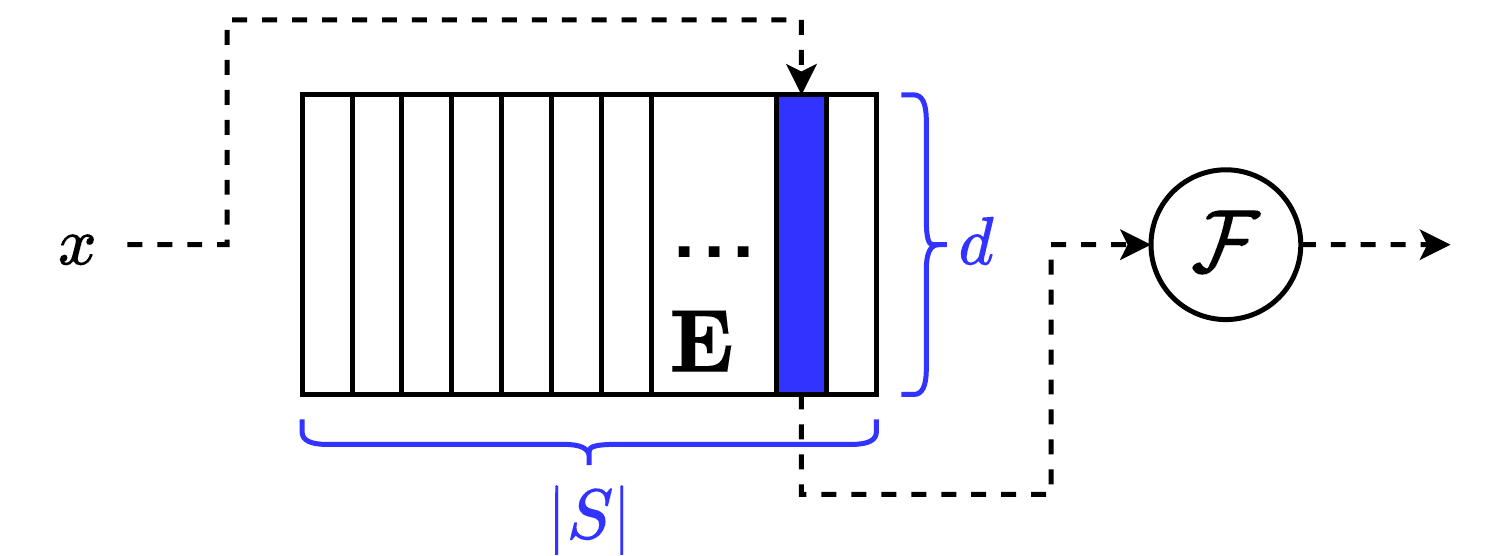}
    \caption{Embedding table $\mathbf{E} \in \mathbb{R}^{|S|{\times}d}$. $x \in \{0,\ldots, |S|-1\}$ is a categorical feature represented by a one-hot vector. $\mathcal{F}$ is a universal function approximator typically parameterized by a neural network.}
    \label{fig:embedding-table}
    \vspace{-0.6cm}
\end{figure}

 \vspace{-0.2cm}
\begin{defn}[Consistent Memory Sharing]
The Fraction of Consistently Shared Memory ($\fcsm$) between two embeddings of size d for values $v_1$ and $v_2$  from $\emb$ under allocation $\mathcal{A}$ is defined as 
\begin{align*}
    f_\mathcal{A}(v_1, v_2) = \frac{1}{d} \ip{\mathcal{A}(v_1)}{\mathcal{A}(v_2)}_\textup{F}
\end{align*}
where $\ip{.}{.}_\textup{F}$ is the Frobenius inner product.
\end{defn}
\vspace{-0.2cm}

We can describe various allocation schemes in terms of the allocation function. For example, $\alloc_{full}$ describes full embedding which is possible when $m { >}(|S|d)$. $\mathcal{A}_h$ describes the na\"{i}ve hashing trick (hash function $h$) based allocation (see section \ref{sec:related_work} for details) which works with any memory budget. Specifically, $\forall v \in S$ and $i \in \{0,..,d{-}1\}$, we have the following.
\begin{align*}
    &\alloc_{full}(v)[i,j] = 1, \quad \textrm{ iff } j=mv+i. \\
    &\alloc_{h}(v)[i,j] = 1, \quad\quad \textrm{ iff }  j=h(v,i).
\end{align*}
It is easy to verify that for every pair of values $v_1$ and $v_2$ $f_{\mathcal{A}_{full}}(v_1,v_2)= 0$  whereas $f_{\mathcal{A}_{h}}(v_1,v_2)= 0$ is random variable with expected value $1/m$. We now define the notation we will use throughout the paper.

\vspace{-0.2cm}
\begin{defn}[General Memory Allocation (GMA)]
\label{def:gma}
Let $\temb$ be the true embedding table which encodes the similarity structure $\tsim$. Let $\emb$ be an embedding table recovered from $\mathcal{M}$ with budget $m$ under allocation $\mathcal{A}$  and $\psim$ be the semantic similarity encoded by $\emb$ Let both similarities be encoded by the kernel function $\tkern{}{}$. We will refer to this as  General Memory Allocation (GMA) setup
\end{defn}
\vspace{-0.2cm}

\textbf{Thought experiment}: We can define the problem of optimal allocation under memory budget when $(m < |S| d)$, under \textit{GMA} setup as
\vspace{-0.1cm}
\begin{align*}
    \argmin_{\alloc}\min_{\mem} \ \zeta(\tsim, \psim)
\vspace{-0.2cm}
\end{align*}
where $\zeta$ is a metric on $\real^{|S|\times|S|}$ (e.g., Euclidean). In order to incorporate the learning of $\mathcal{M}$, we pose the problem as finding $\mathcal{A}$ with best possible associated $\mathcal{M}$ considering that if we choose $\mathcal{A}$ beforehand and then learn $\mathcal{M}$, the learning process will choose the best $\mathcal{M}$. Solving this exact problem appears to be hard. Instead, let us think about an allocation scheme for which we have some evidence of the existence of suitable $\mathcal{M}$. Let us consider a $\mathcal{M}$ with each element independently initialized using $\mathrm{Bernoulli}(0.5, \{-1,+1\})$. If we choose an allocation with constraints based on similarity structure $\tsim$ as
\begin{align*}
    \fcsm(v_1,v_2) = \tsim[v_1,v_2] \quad \forall v_1,v_2 \in S
\end{align*}
then one can verify (as we will show in Theorem \ref{thm:thm_2}) that under this random initialization of memory, pairwise cosine similarity of embeddings of any two values $v_1$ and $v_2$ retrieved from $\mem$ via $\alloc$, denoted as $C_s(v_1, v_2)$, is a random variable with expectation $\mathbb{E}(C_s(v_1,v_2)) = \tsim[v_1,v_2]$ and variance Var$(C_s(v_1, v_2)) \propto \frac{1}{d}$. Hence, this provides evidence that for a semantic similarity based shared allocation, $\alloc_{\tsim}$, there is an assignment to $\mem$, which can produce reasonably small $\zeta( \tsim, \psim)$.  We interpret the $\pm 1$ assignment to each element of embedding as membership to particular concepts, and the overlap in membership determines the similarity. Following this insight, we formally define the semantically constrained memory problem allocation (SCMA) for embeddings as follows.

\vspace{-0.2cm}
\begin{defn}[SCMA Problem]
Under the GMA setup (see Def. \ref{def:gma}), Semantically Constrained Memory Allocation (SCMA) is a problem to find allocation $\mathcal{A}$, under the constraints that for every pair $i,j \in S$, we have $\fcsm(i,j) = \tsim[i,j]$.
\end{defn}
\vspace{-0.2cm}

SCMA problem can be posed as a mixed integer programming (MIP) problem with $\mathcal{O}(|S|^2)$ constraints appearing from similarity constraints along with $\mathcal{O}(|S|d)$ sparsity constraint that the allocation matrix demands. Applying MIP to solve the SCMA problem has the following drawbacks: 1) MIP is computationally intractable for large-scale problems; 2) The solution of MIP needs to be stored explicitly so it is memory expensive; and 3) In case of addition of new values to categorical features, the problem needs to be repeatedly solved. Due to the difficulty of SCMA, in this paper, we formulate and solve the problem using a randomized approach which we term randomized SCMA (RSCMA). We define RSCMA as follows.
\vspace{-0.2cm}
\begin{defn}[RSCMA Problem]
Under the GMA setup (see Def. \ref{def:gma}), the Randomized Semantically Constrained Memory Allocation (RSCMA) is a problem is to find allocation $\mathcal{A}$ under the constraints that for every pair $i,j \in \{0,...,|S|-1\}$, we have 
\begin{align*}
    |f_{\mathcal{A}}(i,j) - \phi(i,j) | \leq \epsilon
\end{align*}
with probability $(1-\delta)$ for some small $\epsilon,\delta > 0$.
\end{defn}
\vspace{-0.2cm}

Compared to applying MIP solvers to the SCMA problem, our approach to RSCMA has the following advantages: 1) LSH based solution is cheaper to compute, 2) storage cost is cheap, and 3) the addition of new values does not require resolving the problem. Although the exact similarity between embeddings is not known a priori, a notion of similarity exists in the data. For example, term-document establishes a similarity between terms based on Jaccard similarity on term-document vectors. We use this similarity to impose a structure on the allocation of memory. Namely, we leverage locality sensitive hashing (LSH) to solve the RSCMA problem. 


\vspace{-0.2cm}
\section{Hashing Schemes}
\vspace{-0.2cm}

We describe important hashing schemes from randomized data structures that will be used in our solution to RSCMA.
\vspace{-0.2cm}
\subsection{Universal Hashing}
\vspace{-0.2cm}

Consider the problem of mapping each element of a large universe $U$ to a smaller range $\{0,\ldots,r{-}1\}$. A family of hash functions $\mathcal{H} = \{h : U \rightarrow \{0,\ldots,r{-}1\} \}$ is $k$ universal if for any $k$ distinct elements $(x_1,\ldots,x_k) \in U$ and any $k$ indices $(\beta_1,\ldots,\beta_k) \in \{0,\ldots,r{-}1\}$, we have $\text{Pr}_{h \in \mathcal{H}}((h(x_1) = \beta_1) \wedge \ldots \wedge (h(x_k) = \beta_k) = 1/r^k$~\cite{univ-hash}. We utilize the following $k$-universal hash function family. 
\begin{align*}
\mathcal{H} = \{ h(x) &= (a_0 + \Sigma_{i=1}^{k-1} a_i x^i ) \, \text{\%} \, P \ \text{\%} \, r \\ 
&| \ a_0 \in \{0,...,P{-}1\}, \ a_i \in \{1,...,P{-}1\} \},
\end{align*}
where $P$ is a large prime number. An instance of hash function drawn from this family is stored by using only $k$ integers $\{a_i\}_{i=0}^{k-1}$.
\vspace{-0.2cm}
\subsection{Locality Sensitive Hashing}\label{sec:LSH} 
\vspace{-0.2cm}
Locality sensitive hashing (LSH) \cite{indyk1998approximate} is a popular tool used in approximate near neighbour search. Consider a function $l : U \rightarrow \{0,...,r{-}1\}$. If $l$ is drawn randomly from a LSH family, say $\mathcal{L}$, then the probability that two elements $x$ and $y$ share same hash value (probability of collision $p_c$) is dependent on a defined similarity measure, $Sim$, between $a$ and $b$. Specifically,
\begin{align*}
    \mathrm{Pr}_{l \in \mathcal{L}}(l(x) == l(y)) \propto Sim(x, y).
\end{align*}
This probability of collision defines a kernel function $\tkern{x}{y}$, which is bounded $0 \leq \tkern{x}{y} \leq 1$, symmetric $\tkern{x}{y} = \tkern{y}{x}$, and reflective $\tkern{x}{x}=1$. We can create multiple LSH families parameterized by $k$, $\mathcal{L}_k$, from a given LSH family $\mathcal{L}$ by using $k$-independently drawn functions from $\mathcal{L}$. Let $\{l_i\}_{i=1}^k$ be $k$ independently drawn functions from $\mathcal{L}$. The new LSH function $\psi \in \mathcal{L}_k$ and its kernel function is defined as 
\begin{align*}
    \psi(x) = (l_1(x), l_2(x), ... l_k(x)) ; \quad  \phi(x,y)_{\mathcal{L}_k} = \phi(x,y)_{\mathcal{L}}^k.
\end{align*}
We call parameter $k$, the {\em power} of LSH functions. The range of certain LSH functions, particularly functions with large power, can be extremely large and needs rehashing into a range, say $\{0,..,r{-}1\}$. This is generally achieved using additional universal hash functions, say $h$ with range $r$. Let the rehashed version of the function $\mathcal{L}$ be denoted by $\mathcal{L}_r$. Then, the kernel of this rehashed LSH is
\begin{align*}
    l_r(x) = h (l(x)) ; \ \  \phi(x,y)_{\mathcal{L}_r} = \phi(x,y)_{\mathcal{L}} + \frac{1- \phi(x,y)_{\mathcal{L}}}{r}.
\end{align*}
\vspace{-0.2cm}
\subsection{Minwise Hashing}
\vspace{-0.2cm}

The minwise hash function is a LSH function that take sets as inputs. The minwise family is as defined below.
\begin{align*}
&\mathcal{L}_{mh} = \{l_\pi | \pi : U \rightarrow U, \pi \text{ is a permutation}\}, \\
&l_\pi(A) = \text{min} (\{\pi(x) | x \in A \}) \quad \textrm{where } A \subseteq U
\end{align*}
For a particular function, $l_\pi$, the hash value of $A$ is the minimum of the permuted values of elements in $A$. As it turns out, the kernel function defined by the collision probability of the minwise hash family is the Jaccard Similarity ($J$).  $J$ measures the similarity between two given sets $A$ and $B$ as $J(A,B)=|A \cap B|/|A \cup B|$. It is easy to check that 
\begin{align*}
    \phi(A,B)_{\mathcal{L}_{mh}} = J(A,B).
\end{align*}

\begin{figure}
    \centering
    \includegraphics[trim = 0 0 0 0, clip, scale=0.4]{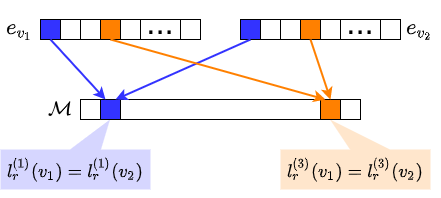}
    \caption{shows consistent sharing of 2 locations when using LMA}
    \label{fig:vis_lma}
    \vspace{-0.5cm}
\end{figure}

\vspace{-0.2cm}
\section{LSH based Memory Allocation(LMA): Solution to RSCMA}\label{sec:lma}
\vspace{-0.2cm}
Consider the standard GMA setup (see Def. \ref{def:gma}) with the semantic structure $\tsim$ defined by a LSH kernel $\tkern{}{}$
We provide an LSH based memory allocation (LMA) solution to this RSCMA. Let the LSH family corresponding to this kernel be $\mathcal{L}$. As defined in Section \ref{sec:LSH}, the probability of collision of values $v_1$ and $v_2$ for corresponding LSH function $l$ and rehashed LSH function $l_r$ can be written as 
\begin{align*}
    &\mathrm{Pr}_{l \in \mathcal{L}}(l(v_1) == l(v_2)) = \phi(v_1, v_2), \\
    &\mathrm{Pr}_{l \in \mathcal{L}}(l_r(v_1) == l_r(v_2)) = \phi(v_1, v_2) + \frac{1-\phi(v_1, v_2)}{m}.
\end{align*}
We use $d$ independently drawn LSH functions $\{l^{(i)}\}_{i=1}^d$. The LMA solution defines the allocation $\mathcal{A}_L$ as
\begin{align*}
    \mathcal{A}_L(v)[i,:] = \textrm{one-hot}(l^{(i)}_r(v)) \quad \forall v \in S,
\end{align*}
where $\textrm{one-hot} : \{0,..m-1\} \rightarrow \{0,1\}^m$ such that for any arbitrary $i \in \{0,..,m-1\}$, $i\neq j$, $\textrm{one-hot}(i)[j] = 0$ and $\textrm{one-hot}(i)[i] = 1$. LMA scheme is illustrated in Figure~\ref{fig:vis_lma}.

In the following theorem, we prove that LMA with the allocation defined by $\mathcal{A}_L$ indeed solves the RSCMA problem.  The proof of this theorem is present in the Supplementary.

\begin{figure}
\begin{subfigure}{.48\linewidth}
  \centering
  \includegraphics[width=.98\linewidth]{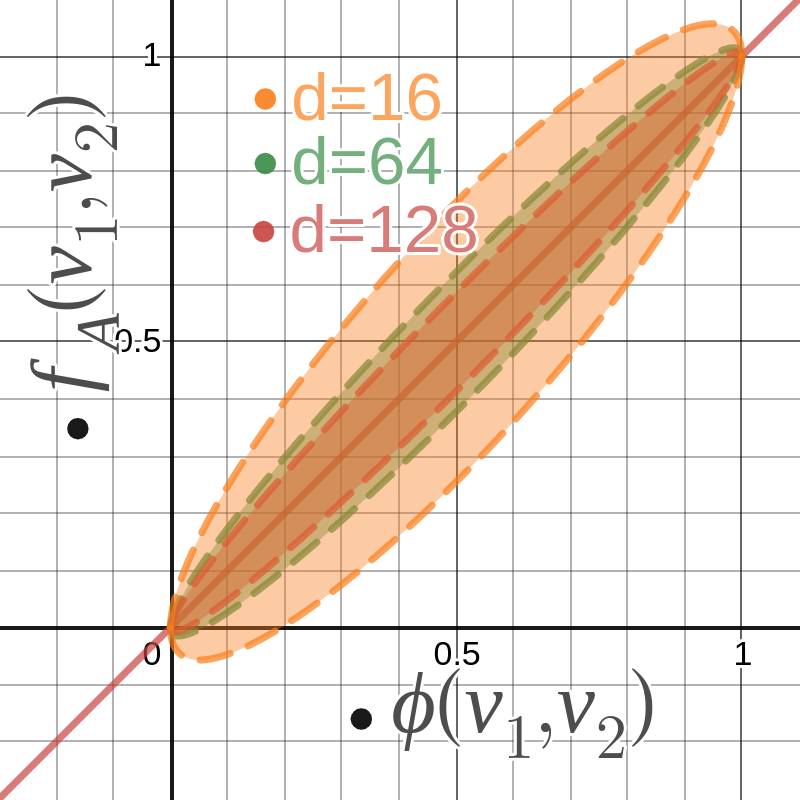}
  \caption{$f_{A_L}(v_1,v_2)$ vs $\phi(v_1, v_2)$}
  \label{fig:sfig1}
\end{subfigure}
\begin{subfigure}{.48\linewidth}
  \centering
  \includegraphics[width=.98\linewidth]{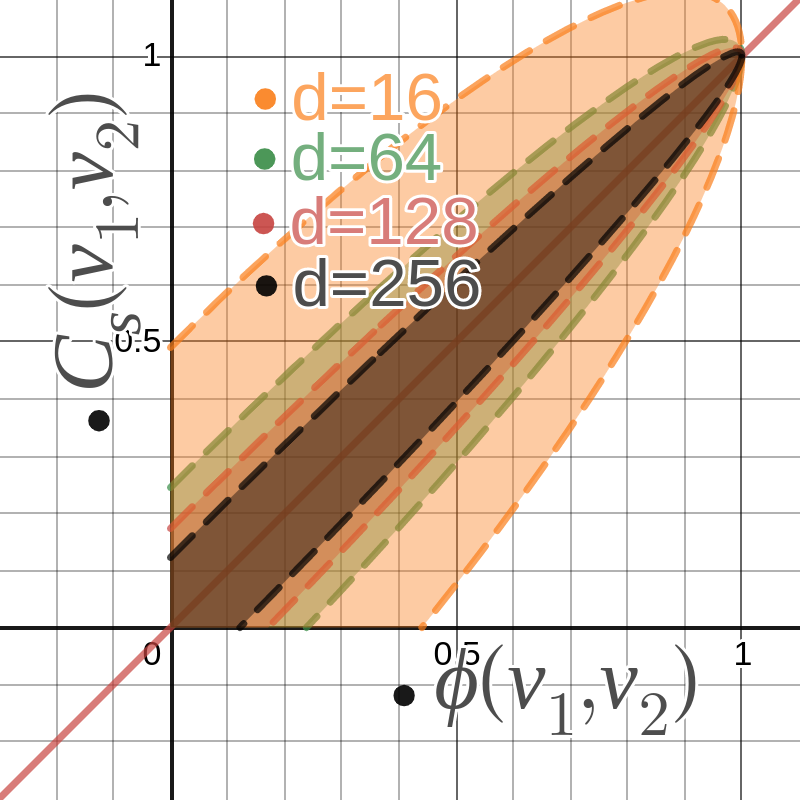}
  \caption{$C_s(v_1,v_2)$ vs $\phi(v_1, v_2)$}
  \label{fig:sf2}
\end{subfigure}
\caption{$\mu \mp 1.96 \sigma $ regions for different values of $d$ as per theorems 1 and 2. For $\hat{f}_{A_L}$ this is 95\% confidence intervals. For $C_s$, this is very close to 95\% confidence interval when m is large }
\label{fig:confidence-intervals}
\vspace{-0.5cm}
\end{figure}

\vspace{-0.2cm}
\begin{thm}[LMA solves RSCMA]\label{thm:thm_lma_rscma}
Under the GMA setup (see def \ref{def:gma}), for any two values $v_1$ and $v_2$, the fraction of consistently shared memory $f_{\mathcal{A}_L}$ as per allocation $\mathcal{A}_L$ proposed by LMA is a random variable with distribution,
\begin{align*}
    &\mathbb{E}(f_{A_L}(v_1, v_2)) =\Gamma =  \phi(v_1, v_2) + \frac{1-\phi(v_1,v_2)}{m}, \\
    &\mathbb{V}(f_{A_L}(v_1, v_2)) = \frac{\Gamma(1-\Gamma)}{d}, \\
    & \mathrm{Pr}\left( |f_{A_L}(v_1, v_2) {-} \phi(v_1, v_2)| > \eta  \Gamma + \frac{1 {-} \phi(v_1,v_2)}{m} \right) \\
    & \quad \quad \quad \leq 2 \exp\left\{\frac{-d \Gamma \eta^2}{3}\right\},
\end{align*}
for all $\eta > 0$. Hence, LMA solves RSCMA with $\epsilon=\eta  \Gamma + \frac{1 {-} \phi(v_1,v_2)}{m}$ and $\delta=2 \exp\left\{\frac{-d \Gamma \eta^2}{3}\right\}$.
\end{thm}
\textbf{Proof sketch}: Proof consists of analyzing  the random variable for the fraction $f_{A_L}(v_1, v_2)$ and applying Chernoff concentration inequality to obtain the tail bounds.\newline
\textbf{Interpretation}:
A reasonable memory $\mathcal{M}$ of 10Mb would have $|\mathcal{M}| > 10^6$. Hence for all practical purposes, we can ignore the $1/m$ terms above. Then, the consistently shared fraction has expected value $\phi(v_1, v_2)$ and variance that is proportional to 1/$d$. A good way to visualize the fact that LMA indeed gives a solution to RSCMA problem is to see the 95\% confidence interval of the fraction, $f_{A_L}(v_1, v_2)$, against the value of $\phi(v_1, v_2)$ as shown in Figure \ref{fig:confidence-intervals}. The parameter $\eta$ is a standard parameter that controls the trade-off between error ($\epsilon)$ and confidence $(1-\delta)$. We can choose $\eta$ very small to reduce error, but then we lose confidence, and instead if we choose $\eta$ large enough to reduce $\delta$, hence increase confidence, then we have more error. 

Next, we prove that if we use LMA to solve RSCMA, we indeed provide an allocation, for which there is an assignment of values to $\mem$ which can lead to an embedding table E, whose associated similarity $\psim$ as measured by cosine similarity is closely distributed around $\tsim$ and hence gives smaller $\zeta (\psim, \tsim)$ (notation as introduced in section \ref{sec:scma}).

\vspace{-0.2cm}
\begin{thm}[Existence of $\mathcal{M}$ with LMA for $\mathbb{S}^*$]
Under the GMA setup (see Def. \ref{def:gma}), let us initialize each element of $\mem$ independently from a $\mathrm{Bernoulli}(0.5,\{-1,+1\})$. Then, the embedding table $\emb$ generated via LMA on this memory, has , for every pair of values $v_1$ and $v_2$, the cosine similarity $C_s(\emb[v_1,:], \emb[v_2,:])$, denoted by $C_s(v_1, v_2)$  is distributed as
\begin{align*}
    &\mathbb{E}(C_s(v_1,v_2)) = \Gamma = \phi(v_1, v_2) + \frac{1 - \phi(v_1, v_2)}{m}, \\
    &\mathbb{V}(C_s(v_1,v_2)) = \frac{1 - \Gamma^2}{d} + \frac{2(1-\Gamma)(d-1)}{dm^2}, \\
    &\mathrm{Pr} \left( |C_s(v_1,v_2) - \phi(v_1, v_2) | \geq \eta \Gamma + \frac{1-\phi(v_1, v_2)}{m} \right)\\
    & \quad \quad \quad \quad \leq \frac{1-\Gamma^2}{d \eta^2 \Gamma^2} \textrm{ for any } \eta > 0.
\end{align*}
\label{thm:thm_2}
\end{thm}
\textbf{Proof sketch}: Proof consists of analyzing the random variable for the cosine similarity $C_s(E[v_1,:], E[v_2,:])$ and applying Chebyshev's concentration inequality to obtain the tail bounds.\newline
\textbf{Interpretation}: We can ignore $1/m$ for any reasonably large memory. Then, the expected value of cosine similarity is exactly $\tkern{v_1}{v_2}$ and it is closely distributed around it. Chebyshev's inequality gives a looser bound and that is apparent from the probable region shown in Figure \ref{fig:confidence-intervals}. In this formulation, again $\eta$ is the parameter controlling error and confidence. Theorem~\ref{thm:thm_2} shows that if we have such a andomly initialized memory, then LMA will lead to intended similarities in approximation.

\vspace{-0.2cm}
\subsection{LMA: Dynamic Solution to RSCMA}
\vspace{-0.2cm}

Unlike any static solution (eg. MIP) to SCMA, LMA solution to RSCMA is highly relevant in a real-world setting. The addition of new features values to datasets is generally frequent, particularly in recommendation data. Any static solution to SCMA will need re-computation every time a value is added. LSH based LMA solution is unaffected by this and can gracefully incorporate new values. 

\vspace{-0.2cm}
\section{LMA for Recommendation Systems
}
\vspace{-0.2cm}
Let us now consider the application of LMA for embeddings of categorical values in ML recommendation systems. DLRM can be used as a running example. However, our approach applies to any system that uses embedding tables. When learning from data in practical settings, $\tsim$ is often unknown as $\temb$ is not known. However, we can use the Jaccard similarity between pairs of values and use the similarity structure to obtain a proxy for $\tsim$. We compute the Jaccard similarity as follows. Let $D_v$ be the set of all sample ids in data in which the value $v$ for some categorical feature appears (e.g., a row in a term-document matrix). Then, we can define the similarity $\tsim$ as
\begin{equation*}
    \tsim[v_1,v_2] = J(D_{v_1}, D_{v_2}).
\end{equation*}
The resulting kernel is the Jaccard kernel, which is an LSH kernel with the minwise LSH function. To hash a value, say $v$, with the minwise hash function, we will use the corresponding set $D_v$.  We can then use the general LMA setup described in Section $\ref{sec:lma}$ to adjust our training algorithms. 

\textbf{Size of data required:}
It may appear that storing the data for minhash computations would diminish memory savings. However, recall that we need data only to obtain a good estimate of Jaccard. We find that the actual data required, $D'$, is significantly less than the total data. Namely,$|D'|\ll|S|d$. Therefore, to formalize, we bound the number of non-zeros required for Jaccard computation.
\vspace{-0.2cm}
\begin{thm}[$D'$ required is small] \label{thm:thm_3}
Assuming a uniform sparsity of each value, s, the Jaccard Similarity of two values x and y,say $J=J(D_x, D_y)$ when estimated from a i.i.d subsample $D' \subseteq \mathcal{D}$, $|D'| = n$ is distributed around J as follows.
\begin{equation*}
    | \mathbb{E}(\hat{J}) - J | \leq \epsilon J,
\end{equation*}
\begin{equation*}
    |\mathbb{V}(\hat{J}) - A| \leq 2\epsilon(A+2J^2),
\end{equation*}
where $A =\frac{J}{2ns} (1+J -2sJ)$ with probability $(1-\delta)$ where $\delta = \frac{1+J - 2s}{2ns\epsilon^2}$.
\end{thm}
\textbf{Interpretation:} The bound given by Theorem~\ref{thm:thm_3} is loose due to the approximations done to analyze this random variable (see Supplemental). Nonetheless, it can be seen that for a given value of $\delta$, we can control $\epsilon$ and $A$ through $ns$, which is the number of non-zeros per value. As $ns$ increases, both the variance and deviation of the expected value from $J$ decrease rapidly. In practice, Section~\ref{sec:experiments} shows we only need around 100K samples for the Criteo dataset out of 4.5M samples.
\begin{figure}
    \centering
    \includegraphics[scale=0.26]{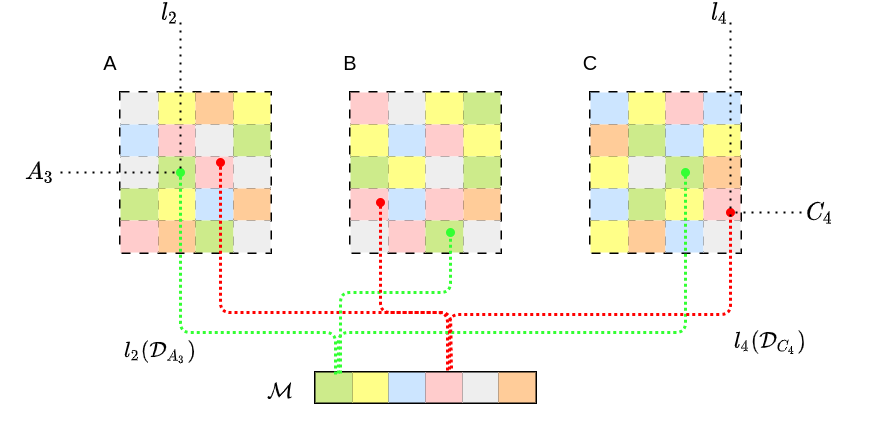}
    \caption{LMA for DLRM model using common memory across the tables}
    \label{fig:ssmdlrm}
    \vspace{-0.5cm}
\end{figure}
Below we discuss a few considerations in relation to LMA applied to DLRM model.\newline
\textbf{Memory Comparison:}
The size of the memory used by full embeddings is $\mathcal{O}(|S|d)$. The linear dependence on $d$ and the typical very large size of set $S$ makes it difficult to train and deploy large dimensional embedding models. LMA solutions to RSCMA make it possible to simulate the embeddings of size $\mathcal{O}(|S|d)$ using any memory budget $m$. The memory cost with LMA procedure comprises of: 1) Memory budget: $\mathcal{O}(m)$ $m=|\mem|$; 2) Cost of storing LSH function: $\mathcal{O}(k d + k')$ required to store $d$ minhash functions with power $k$ and $k'$-universal hash function for rehashing. Generally these values are very small compared to $\mathcal{O}(|S|d)$ and can be ignored; and 3) The size of Data $D'$ stored and used for minhash functions. Generally, size of $D'$ is much smaller than $(|S|d)$. For example in Criteo, the subsample we used had around 3M integers (when using 125K samples) as compared to the range of  50M-540M floating parameters of the models we train. We empirically analyze the effect on various sizes of $D'$ in the experimental section. This requirement of smaller $D'$ is also an effect of the way we handle very sparse features, which is discussed next. To summarize, the memory cost of LMA is $\mathcal{O}(m + kd + k' + |D'|) \approx \mathcal{O}(m)$. \newline
\textbf{Handling very sparse features:} For very sparse features, the embedding quality does not significantly affect the accuracy of the model \cite{MDTrick19}. Also, it is difficult to get a good estimate of Jaccard similarity for this using a small subsample. Due to these reasons, for very sparse features, we randomly map each element of its embedding into $\mathcal{M}$. Essentially, we revert to $\mathcal{A}_h$ (na\"{i}ve hashing trick) for such values.\newline
\textbf{Common Memory:} We use a single common memory $\mathcal{M}$ across all embedding tables in DLRM. The idea is to fully utilize all similarities to share maximum memory and hence get the best memory utilization.\newline
\textbf{Forward Pass:}
The forward pass requires retrieving embeddings from $\mathcal{M}$. Let $V_{batch}$ be the set of all values in a batch. We collect the set $\{D_v  | v{\in}V_{batch} \}$, apply GPU friendly implementation of $d$-minhashes to it to obtain a matrix of locations, $I \in R^{|V_{batch}| \times d}$. Using this we get $E_{batch} = \mathcal{M}[I]$.\newline
\textbf{Backward Pass:}
The memory $\mathcal{M}$ contains the parameters which are used to construct embeddings. Hence, in the backward pass, the gradient of parameters in $\mathcal{M}$ is computed and these parameters are updated. The exact functional dependence of the result on a parameter in $\mathcal{M}$ is complex as it is implemented via LSH mappings. Auto gradient computation packages of deep learning libraries (e.g., PyTorch and TensorFlow) are used for gradient computation.

\begin{figure*}
\centering
\begin{subfigure}{.3\textwidth}
    \includegraphics[width=1\linewidth]{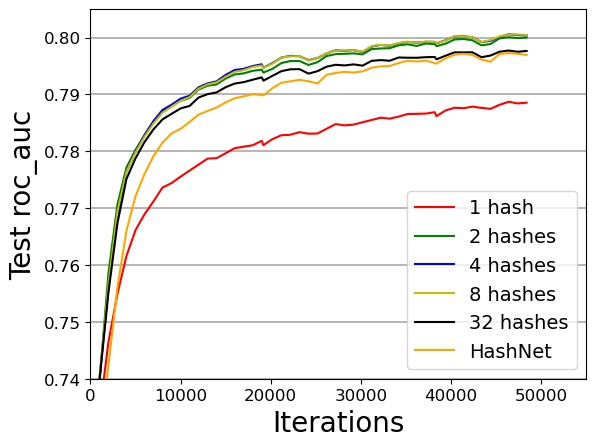}
    \caption{Varying $n_h$ with $\alpha{=}32\times$, $n_s{=}125K$}
    \label{fig:number_of_hash_functions}
\end{subfigure}
\begin{subfigure}{.3\textwidth}
    \includegraphics[width=1\linewidth]{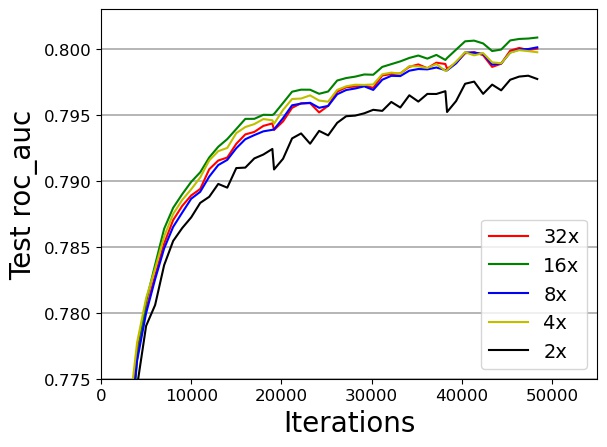}
    \caption{Varying $\alpha$ with $n_h{=}2$, $n_s{=}125K$}
    \label{fig:expansion_rate}
\end{subfigure}
\begin{subfigure}{.3\textwidth}
    \includegraphics[width=1\linewidth]{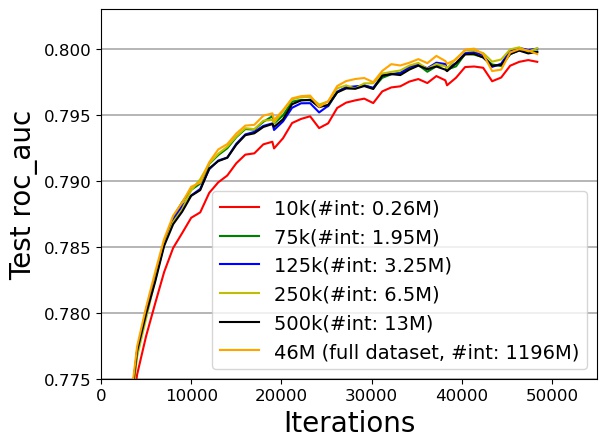}
    \caption{Varying $n_s$ with $n_h=2$, $\alpha{=}32 \times$}
    \label{fig:representation_size}
\end{subfigure}
\vspace{-0.2cm}
\caption{Effect of hyperparameters on performance of LMA-DLRM }
\label{fig:hypexp}
\vspace{-0.5cm}
\end{figure*}

\vspace{-0.2cm}
\section{Related Work}\label{sec:related_work}
\vspace{-0.2cm}

We focus on related works that significantly reduce the size of the embedding matrix for recommendation and click-through prediction systems. Namely, hashing trick~\cite{hashtrick}, compositional embedding using complementary partitions~\cite{QuoRemTrick19}, and mixed dimension embedding~\cite{MDTrick19}.

{\bf Na\"{i}ve hashing trick:} Given the embedding table $\mathbf{E} \in \mathbb{R}^{|S|{\times}d}$, two basic approaches that leverage hashing trick are presented.
\begin{itemize}[leftmargin=*, nosep]
\item Vector-wise (or row-wise): let $\mathbf{\hat{E}} \in \mathbb{R}^{m{\times}d}$ such that $m{\times}d = |\mathcal{M}|$, the memory budget, and $m \ll |S|$ be the reduced size embedding table. We use a hash function $h : \{0,1,\ldots,|S|{-}1\} \rightarrow \{0,1,\ldots,m{-}1\}$ that maps the (row-wise) indices of the embeddings from the full embedding table $\mathbf{E}$ to the reduced embedding table $\mathbf{\hat{E}}$. The size of the embedding table is reduced from $\mathcal{O}(|S|d)$ to $\mathcal{O}(md)$.
\item Element-wise (or entry-wise): let $\mathbf{\hat{E}} \in \mathbb{R}^m$ such that $m=|\mathcal{M}|$. We use a hash function $h : \{(i,j)\}_{i,j=0}^{|S|,d} \rightarrow \{0,1,\ldots,m{-}1\}$  that maps each element $\mathbf{E}_{i,j}$ to an element in $\mathbf{\hat{E}}$. The size of the embedding table is reduced from $\mathcal{O}(|S|d)$ to $\mathcal{O}(m)$~\cite{hashtrick}.
\end{itemize}

{\bf Compositional embedding using complementary partitions:} In the vector-wise hashing trick, multiple embedding vectors are mapped to the same index, which results in loss of information on the unique categorical values and reduction in expressiveness of the model. To overcome this issue, \cite{QuoRemTrick19} proposes to construct {\em complementary partitions} of $\mathbf{E}$, from set theory, and apply {\em compositional operators} on embedding vectors from each partition table to generated unique embeddings. One example of a complementary partition is the so-called quotient-remainder (QR) trick. Two embedding tables $\mathbf{\hat{E}}_1 \in \mathbb{R}^{m{\times}d}$ and $\mathbf{\hat{E}}_2 \in \mathbb{R}^{(|S|{/}m){\times}d}$ are created, and two hash functions $h_1$ and $h_2$ respectively are used for mapping. $h_1$ maps the $i$-th row of $\mathbf{E}$ to the $j$-th row of $\mathbf{\hat{E}}_1$ using the remainder function: $j = i \mod m$. $h_2$ maps the $i$-th row of $\mathbf{E}$ to the $k$-th row of $\mathbf{\hat{E}}_2$ using the function $k = i \sslash m$, where $\sslash$ denotes integer division (quotient). Taking the embeddings $e_j \in \mathbf{\hat{E}}_1$ and $e_k \in \mathbf{\hat{E}}_2$ and applying element-wise multiplication $e_j \odot e_k$ results in a unique embedding vector. The resulting memory complexity is $\mathcal{O}(\frac{|S|}{m}d + md)$. In general, the optimal memory complexity is $\mathcal{O}(k|S|^{1/k}d)$ with $k$ complimentary partitions of sizes $\{m_i \times d \}_{i=1}^k$ such that $|S| \leq \Pi_{i=1}^k m_i$.


{\bf Mixed dimension (MD) embedding:} Frequency of categorical values are often skewed in real-world applications. Instead of a fixed (uniform) embedding dimension for all categories,~\cite{MDTrick19} proposes that embedding dimensions scale according to popularity of categorical values. Namely, more popular values are set to higher dimension embeddings (i.e., allocate more memory) and vice versa. The idea is to create embedding tables, along with a projection matrix, of the form $\{(\mathbf{\hat{E}}_i, \mathbf{\hat{P}}_i)\}_{i=1}^k$, such that $\mathbf{\hat{E}}_i \in \mathbb{R}^{|S_i|{\times}d_i}$ (MD embedding), $\mathbf{\hat{P}}_i \in \mathbb{R}^{d_i{\times}\bar{d}}$ (projection), $\bar{d} \geq d_i$, and $|S|=\sum_{i=1}^k |S_i|$. $k$ and $\mathbf{d}=(d_1,\ldots,d_k)$ are input parameters. One approach proposed to set the parameters is based on a power-law sizing scheme using a meta {\em temperature} parameter $\alpha$. Let $\mathbf{p} = \{p_1,\ldots,p_k\}$, where $p_i$ is a probability value (e.g., $p_i = 1/|S_i|$). Then, $\lambda = \bar{d}||\mathbf{p}||_{\infty}^{-\alpha}$  is the scaling factor and  $\mathbf{d} = \lambda\mathbf{p}^{\alpha}$ is the component-wise exponent. $\alpha=0$ implies uniform dimensions and $\alpha=1$ sets dimensions proportional to popularity.

\textbf{Comparison with LMA:}
Both QR Trick and MD Trick change the embedding design. LMA does not affect embedding design directly. Instead, it solves the abstract problem of RSCMA. We can apply LMA in conjunction with any such embedding design tricks to obtain better memory utilization.

\newcommand{\tabincell}[2]{\begin{tabular}{@{}#1@{}}#2\end{tabular}}

\begin{figure*}\label{fig:main_experiment}
\centering
\begin{subfigure}{.33\textwidth}
    \centering
    \includegraphics[width=.99\linewidth]{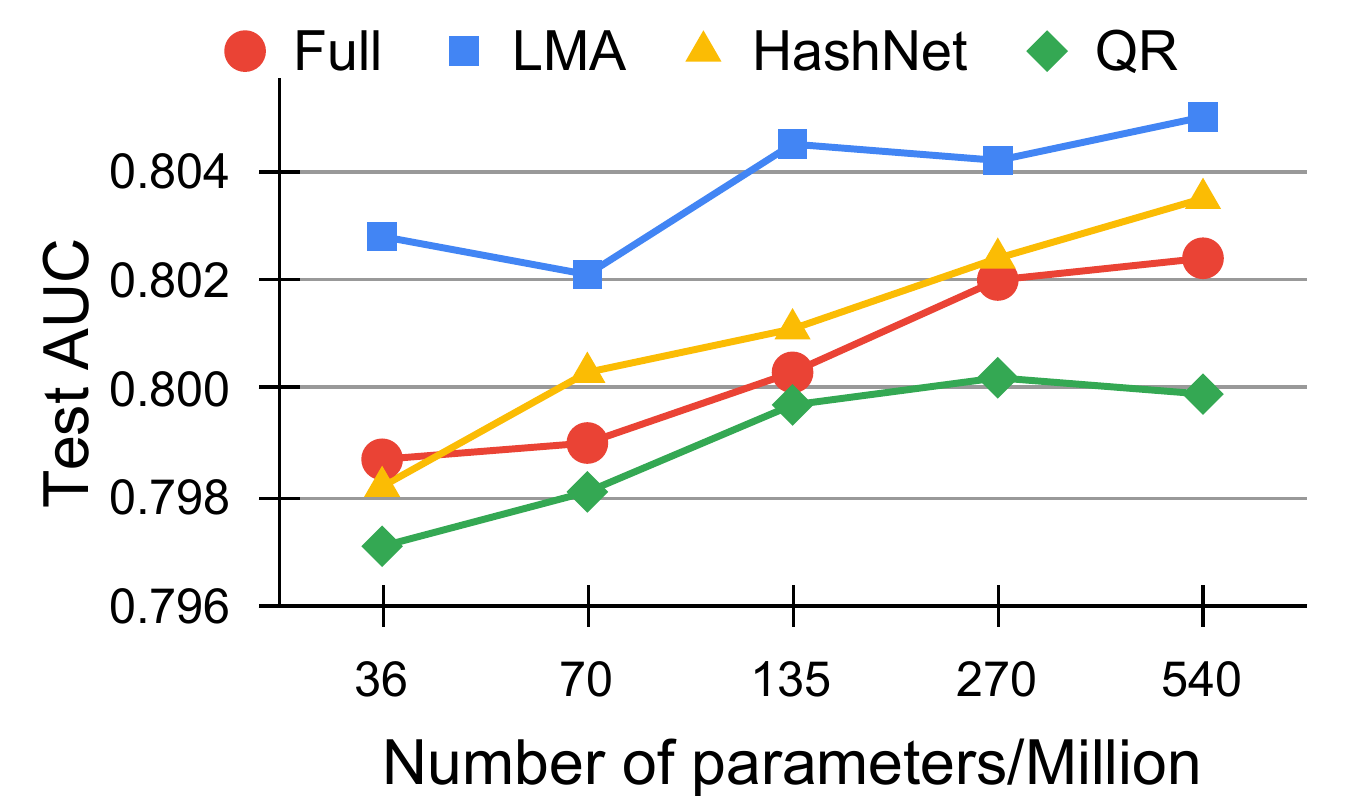}
\end{subfigure}
\begin{subfigure}{.33\textwidth}
    \centering
    \includegraphics[width=.99\linewidth]{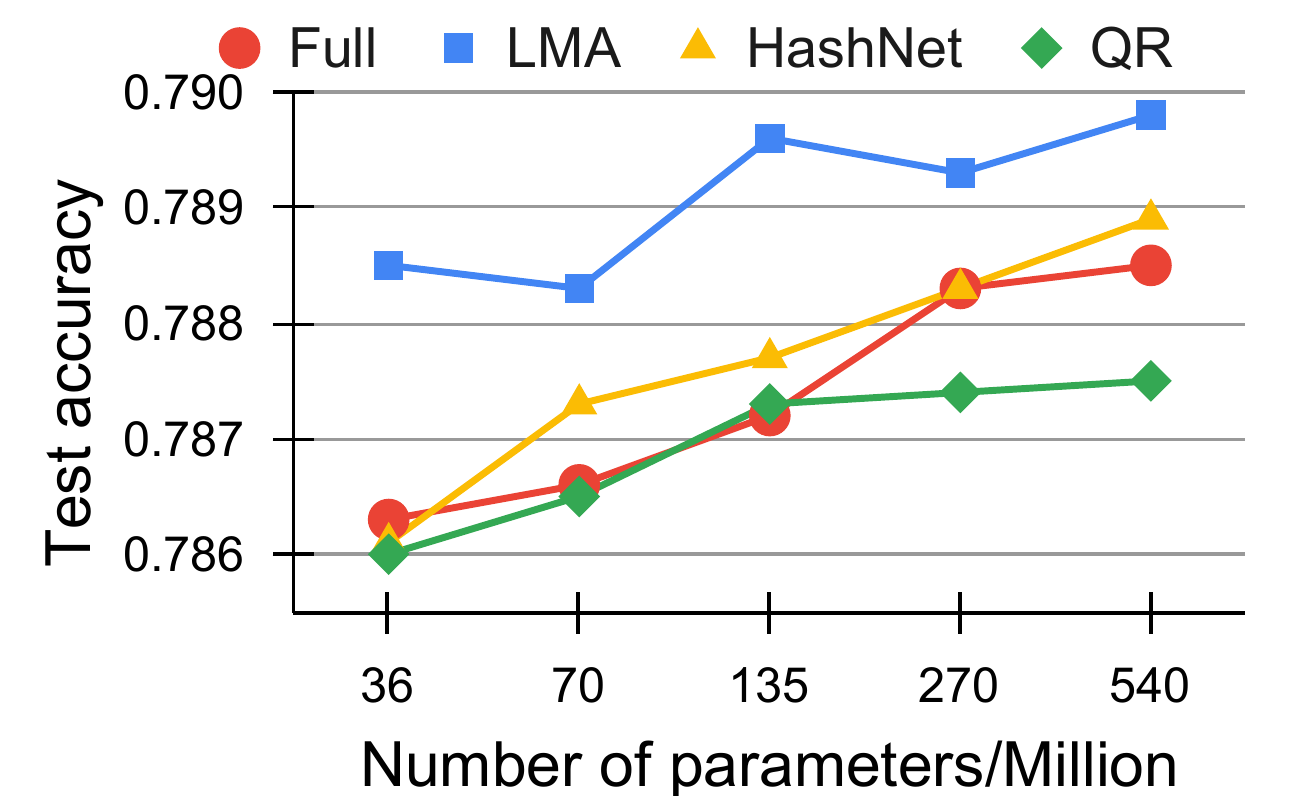}
\end{subfigure}
\begin{subfigure}{.33\textwidth}
    \centering
    \includegraphics[width=.99\linewidth]{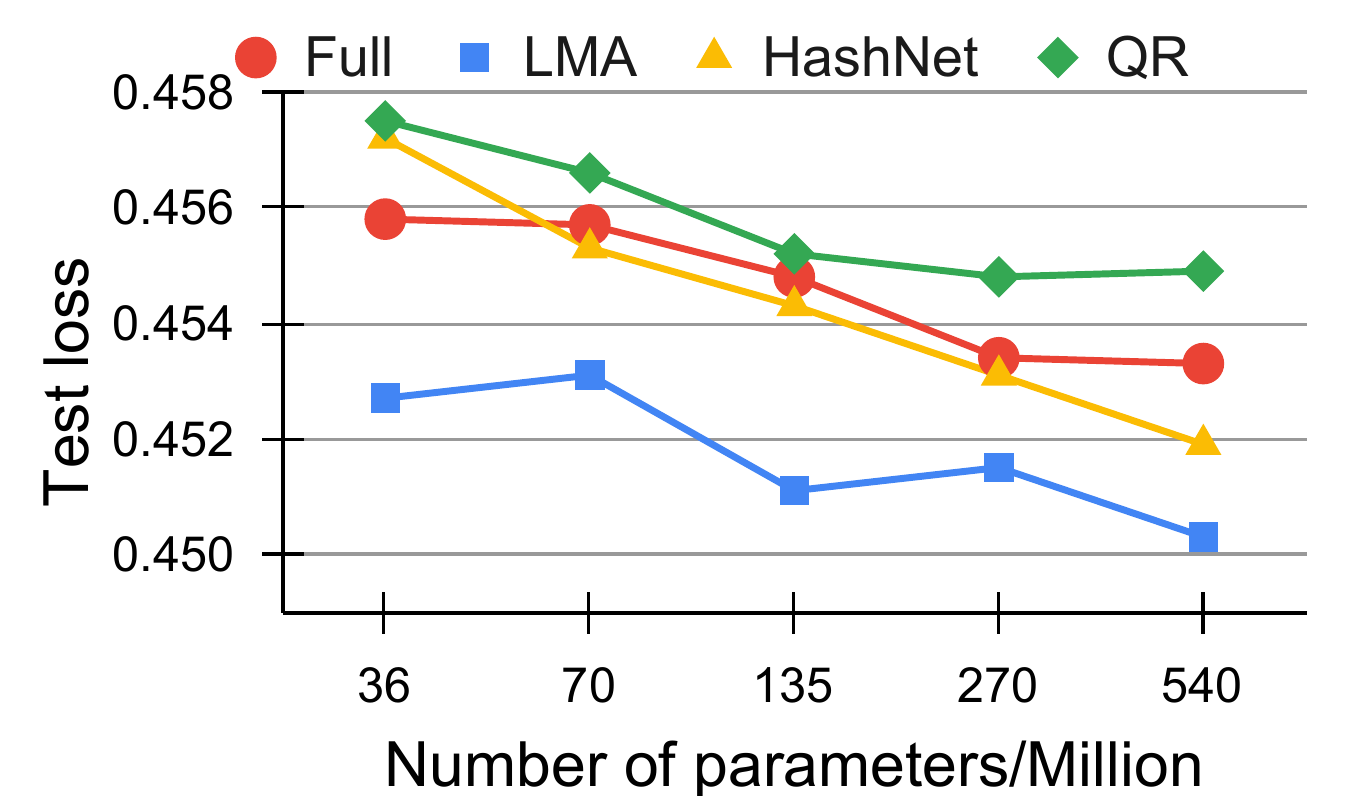}
\end{subfigure}
\begin{subfigure}{.33\textwidth}
    \centering
    \includegraphics[width=.99\linewidth]{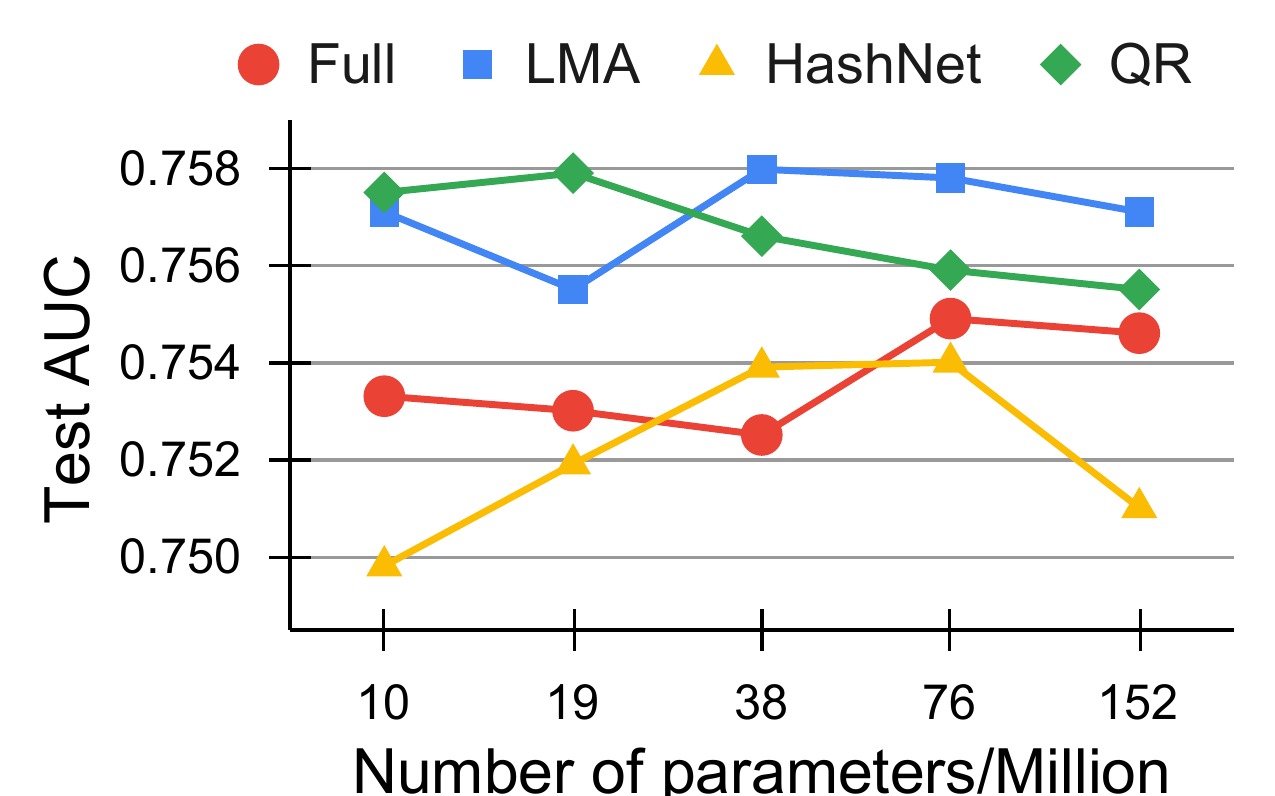}
\end{subfigure}
\begin{subfigure}{.33\textwidth}
    \centering
    \includegraphics[width=.99\linewidth]{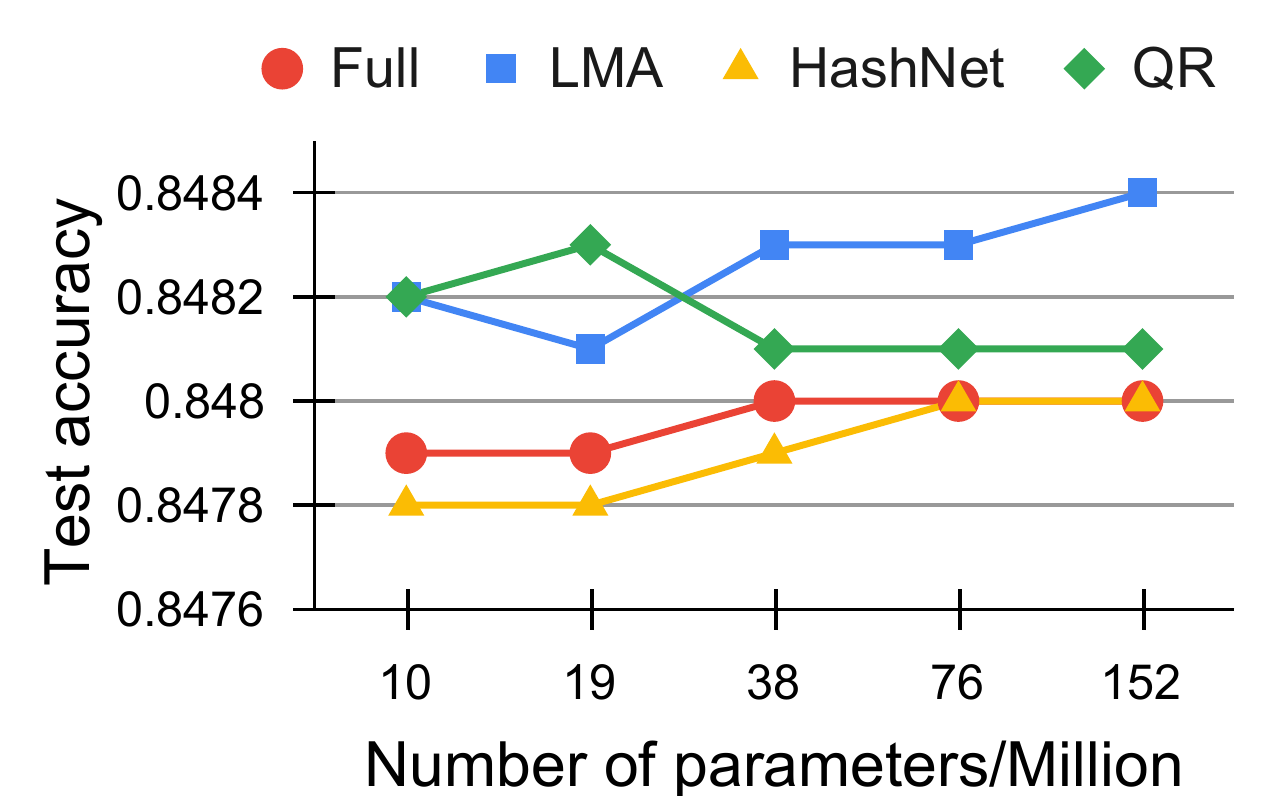}
\end{subfigure}
\begin{subfigure}{.33\textwidth}
    \centering
    \includegraphics[width=.99\linewidth]{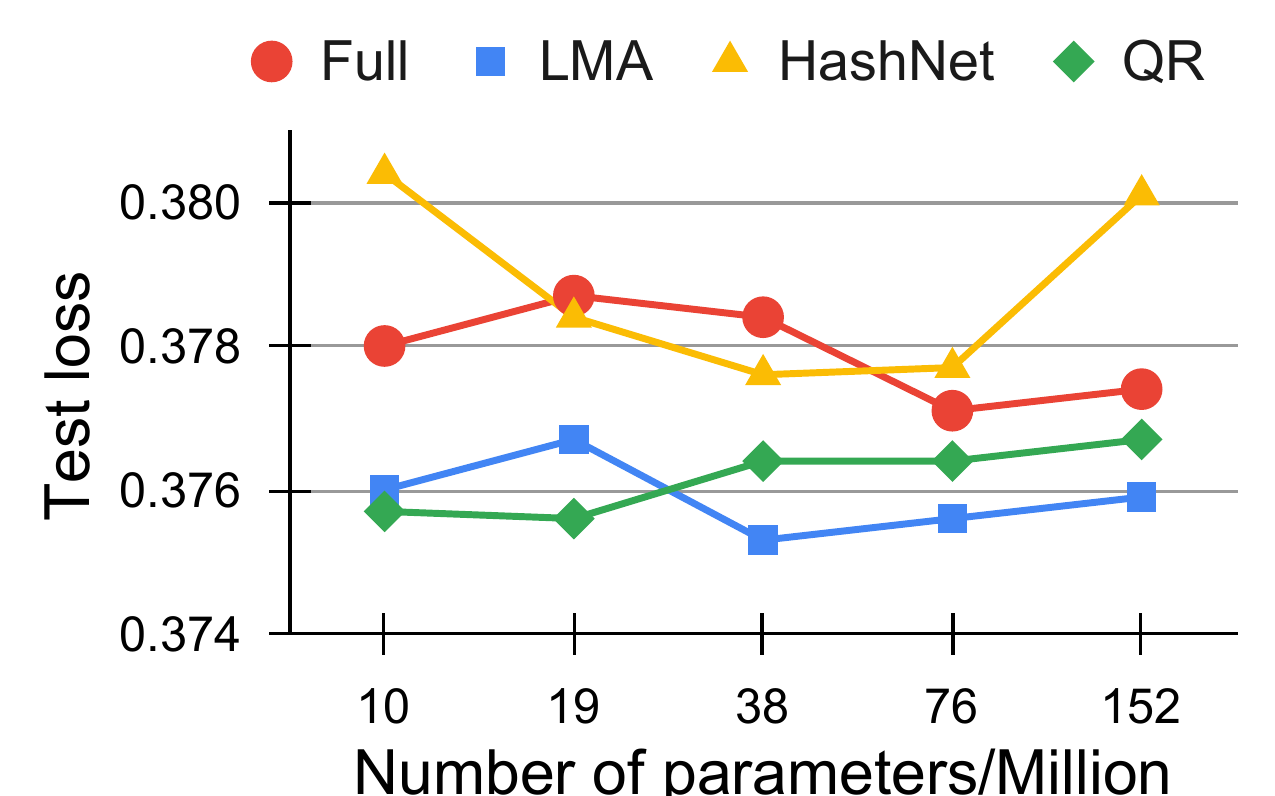}
\end{subfigure}
\caption{AUC, Accuracy, and Loss against memory regimes (number of parameters) on Criteo (\textbf{top}) and Avazu (\textbf{bottom})}
\label{fig:main_experiment}
\vspace{-0.5cm}
\end{figure*}

\vspace{-0.2cm}
\section{Experiments}\label{sec:experiments}
\vspace{-0.2cm}

\textbf{Datasets:}
To evaluate the performance of LMA on DLRM (LMA-DLRM), we use two public click-through rate (CTR) prediction datasets: Criteo and Avazu. Criteo is used in the DLRM paper~\cite{DLRM19} and related works focused on memory-efficient embeddings~\cite{MDTrick19, QuoRemTrick19}. Avazu is a mobile advertisement dataset. A summary of dataset properties is presented in Table~\ref{tab:datasets}. Values represent the number of all the categorical values. The number of values per feature varies a lot : for example, some lower values are 10K and they go as high as 10M.

\begin{table}[]
\centering
\caption{Description of datasets. cat: categorical, int: integer.}
\begin{tabular}{|l|l|l|l|l|}
\hline
       & {\bf \#Samples} & \begin{tabular}[c]{@{}l@{}}{\bf \#Features}\\ {\bf (cat+int)}\end{tabular} & \begin{tabular}[c]{@{}l@{}}{\bf Positive} \\ {\bf rate}\end{tabular} & {\bf \#Values} \\ \hline
Criteo & 46M     & 26 + 13                                                      & 26\%                                                    & 33.76M  \\ \hline
Avazu  & 41M     & 21 + 0                                                       & 17\%                                                    & 9.45M  \\ \hline
\end{tabular}
\label{tab:datasets}
\vspace{-0.6cm}
\end{table}

\textbf{Metrics:} We use loss, accuracy, and ROC-AUC as metrics for our experiments. For imbalanced datasets like these, AUC is a better choice of metric than accuracy. \newline
\textbf{Basic setup:} For all our experiments (LMA-DLRM, baselines), we follow the basic setup (e.g., optimizer parameters, model architecture) suggested by the DLRM paper~\cite{DLRM19} and as implemented in their GitHub page. The batch size is set to 2,048 as we want to run experiments for a larger number of epochs. Also, a higher batch size is preferred for CTR datasets~\cite{zhu2020fuxictr}. We use the same settings for LMA-DLRM as well as all baselines for a fair comparison.

\vspace{-0.2cm}
\subsection{Hyperparameter experiments}\label{sec:hyperparameter_experiment}
\vspace{-0.2cm}
There are three hyperparameters: $n_h$ : power of LSH function see \ref{sec:LSH}, $\alpha$: expansion rate, and $n_s$ : number of samples in $D'$ for LMA-DLRM. The best values for each of these hyperparameters can be chosen via cross-validation. We describe these parameters and qualitatively analyze their effect on performance. The results of changing one parameter while keeping the other two fixed are shown in Figure \ref{fig:hypexp}. We use 270M budget for varying $\alpha$ and 35M budget for others.
\begin{itemize}[leftmargin=*,nosep]
    \item \textbf{Power of LSH ($n_h$)}
    The $n_h$ used per LSH mapping (i.e. \emph{power} in Section \ref{sec:LSH}) controls the probability of collision (i.e., $J^{n_h}$) of corresponding elements of different embeddings as discussed in Section \ref{sec:LSH}. Higher power leads to a lower probability of collision, making the rehashed LSH function behave more like na\"ive hashing trick. Very low power will increase memory sharing and might lead to its under-utilization.  This phenomenon can be observed in Figure \ref{fig:number_of_hash_functions}, where $n_h{=}1$ gives the worst performance. Increasing $n_h$ improves the performance until $n_h=8$. Performance worsens when $n_h=32$ and tends towards the hashing trick performance as expected.
    
    \item {\textbf{Expansion rate ($\alpha$)}}
    LMA-DLRM can simulate embedding tables of any dimension $d$. We define expansion rate as the ratio of simulated memory to actual memory. Using GMA notation, $\alpha = |S|d/|\mem|$. Figure \ref{fig:expansion_rate} shows that 16 works best for memory budget of $\sim$ 270M parameters. So increasing $\alpha$ will not improve the performance indefinitely. 
    \item {\textbf{Samples in $D'$ ($n_s$)}:} Figure \ref{fig:representation_size} shows that the performance boost saturates after the representation size reaches 75k data points, which means that most of the frequently appearing values (v) get decent representations ($D_v$) from a small number of samples. For very sparse values, we revert to element-wise na\"ive hashing trick based mapping. We also show the size of the samples in terms of the number of non-zeros integers to be stored. As compared to 540M parameter networks we train a sample of 125K only requires us to store 3.2M integers.
\end{itemize}
\vspace{-0.2cm}
\subsection{Main Experiment}
\begin{figure}
    \centering
    \begin{subfigure}{.48\linewidth}
    \centering
    \includegraphics[width=0.98\textwidth]{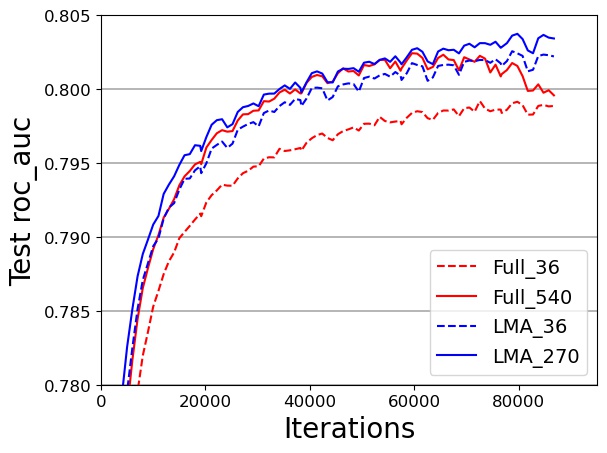}
\end{subfigure}
\begin{subfigure}{.48\linewidth}
    \centering
    \includegraphics[width=0.98\textwidth]{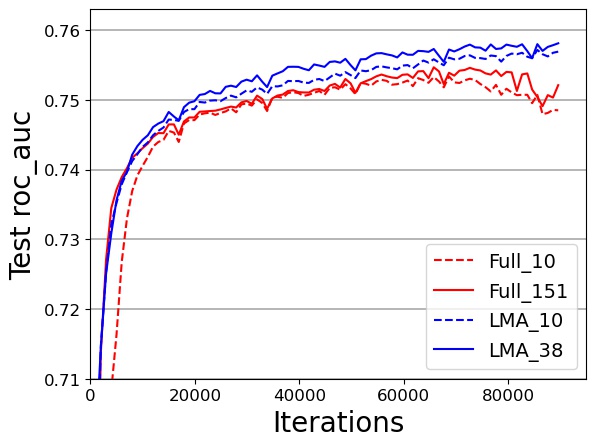}
\end{subfigure}
    \caption{Evolution of test AUC with different parameters on Criteo \textbf{(left)} and Avazu \textbf{(right)} }
    \label{fig:main_evolution}
    \vspace{-0.5cm}
\end{figure}

\vspace{-0.2cm}
We compare LMA-DLRM against full embedding (embedding tables used in DLRM~\cite{DLRM19}), HashedNet~\cite{hashtrick} embedding (na\"{i}ve element-wise hashing trick based), and QR embeddings~\cite{QuoRemTrick19} across different memory budgets. Hyperparameters $n_h {=} 4$, $\alpha {=} 16$, and $n_s{=}125,000$ were used for all LMA-DLRM experiments in this section. For baselines, we use the configurations in their open source code. Training of models was cutoff at 15 epochs. We did not perform extensive hyperparameter tuning. However, hyperparameter tuning can achieve better results for LMA-DLRM. The LMA-DLRM code is attached with supplementary material.\newline
\textbf{Results}:
Figure~\ref{fig:main_experiment} shows AUC, accuracy, and loss against different memory regimes (number of parameters) on both datasets. Figure~\ref{fig:main_evolution} shows the evolution of some models for first 5 epochs. 
\begin{itemize}[leftmargin=*,nosep]
    \item LMA-DLRM outperforms all baselines across different memory regimes include those reported in \cite{DLRM19}, achieving average AUC improvement of \textbf{0.0032} in Criteo and \textbf{0.0034} in Avazu across memory budgets and an average improvement in Accuracy of \textbf{0.0017} on Criteo. Recall that an improvement of 0.001 is significant.
    \item The AUC and accuracy of full embedding model with 540M parameters can be achieved by LMA-DLRM with only 36M parameters ($\mathbf{16}{\boldsymbol{\times}}$ {\bf reduction}) on Criteo. On Avazu, results with 10M parameter LMA model are much better than full embedding model with 150M parameters ($\mathbf{15}{\boldsymbol{\times}}$ {\bf reduction}).
    \item LMA-DLRM achieves best AUC \textbf{0.805} as opposed to the best of full embedding \textbf{0.802} on Criteo. On Avazu, we see improvement of \textbf{0.0025} on best AUCs as well. 
    \item The typical evolution of AUC metric for Criteo and Avazu on models of different sizes for LMA-DLRM and full embedding models clearly supports the better performance of LMA-DLRM over full embeddings.
\end{itemize}

\vspace{-0.2cm}
\section{Conclusion}
\vspace{-0.2cm}
We define two problems namely SCMA (Semantically Constrained Memory Allocation) and Randomized SCMA for efficient utilization of memory in Embedding tables. We propose a neat LSH-based Memory Allocation (LMA) which solves the dynamic version of RSCMA under any memory budget with negligible memory overhead. LMA was applied to an important problem of heavy memory tables in widely used recommendation models and found tremendous success. In this paper, we focus on the memory aspect of LMA. In future work, we would like to benchmark the LMA method for its time efficiency for training and inference for recommendation models.
\newpage
\bibliography{refs-ssm}
\bibliographystyle{unsrt}

\onecolumn
\appendix
\section{LMA solves RSCMA}
Under the GMA setup (see def \ref{def:gma}), for any two values $v_1$ and $v_2$, the fraction of consistently shared memory $f_{\mathcal{A}_L}$ as per allocation $\mathcal{A}_L$ proposed by LMA is a random variable with distribution,
\begin{align*}
    &\mathbb{E}(f_{A_L}(v_1, v_2)) =\Gamma =  \phi(v_1, v_2) + \frac{1-\phi(v_1,v_2)}{m}, \\
    &\mathbb{V}(f_{A_L}(v_1, v_2)) = \frac{\Gamma(1-\Gamma)}{d}, \\
    & \mathrm{Pr}\left( |f_{A_L}(v_1, v_2) {-} \phi(v_1, v_2)| > \eta  \Gamma + \frac{1 {-} \phi(v_1,v_2)}{m} \right) \\
    & \quad \quad \quad \leq 2 \exp\left\{\frac{-d \Gamma \eta^2}{3}\right\},
\end{align*}
for all $\eta > 0$. Hence, LMA solves RSCMA with $\epsilon=\eta  \Gamma + \frac{1 {-} \phi(v_1,v_2)}{m}$ and $\delta=2 \exp\left\{\frac{-d \Gamma \eta^2}{3}\right\}$.

\textbf{Proof sketch}: Proof consists of analyzing  the random variable for the fraction $f_{A_L}(v_1, v_2)$ and applying Chernoff concentration inequality to obtain the tail bounds.\newline
\textbf{Proof}

The probability that a particular location is shared is exactly the probability of collision of the rehashed lsh function $l_r$. (using the notation from section \ref{sec:lma})
\begin{equation}
    \mathrm{Pr}(l_r(x) == l_r(y)) = \phi(x,y) + \frac{(1 - \phi(x, y))}{m}
\end{equation}
We can write the indicator for fraction of consistently shared memory as,
\begin{equation}
    \hat{f_{\mathcal{A}_L}} =  \sum_{i=1}^d \mathcal{I}(l_r^{(i)}(x) == l_r^{(i)}(y))
\end{equation}
  $f_{\mathcal{A}_L}$ is a sum of independent bernoulli varaibles. 
  It is easy to check that the expected value of consistently shared fraction is, 
\begin{equation}
    E(\hat{f_{\mathcal{A}_L}}) = \Gamma =  \phi(x,y) + \frac{(1 - \phi(x, y))}{m}
\end{equation}
\begin{equation}
    V(\hat{f_{\mathcal{A}_L}}) = \frac{\Gamma (1-\Gamma)}{d}
\end{equation}
We can apply the chernoff's bound to get tail bound. Let $\eta > 0$ be any positive real number
\begin{align}
&\mathrm{Pr}\left( |f_{A_L}(v_1, v_2) {-} \Gamma | > \eta  \Gamma  \right)   \leq 2 \exp\left\{\frac{-d \Gamma \eta^2}{3}\right\} \\
&\mathrm{Pr}\left( |f_{A_L}(v_1, v_2) {-} \phi(v_1, v_2)| > \eta  \Gamma + \frac{1 {-} \phi(v_1,v_2)}{m} \right) \leq 2 \exp\left\{\frac{-d \Gamma \eta^2}{3}\right\}    
\end{align}

\section{Existence of $\mathcal{M}$ with LMA for $\mathbb{S}^*$}

Under the GMA setup (see Def. \ref{def:gma}), let us initialize each element of $\mem$ independently from a $\mathrm{Bernoulli}(0.5,\{-1,+1\})$. Then, the embedding table $\emb$ generated via LMA on this memory, has , for every pair of values $v_1$ and $v_2$, the cosine similarity $C_s(\emb[v_1,:], \emb[v_2,:])$, denoted by $C_s(v_1, v_2)$  is distributed as
\begin{align*}
    &\mathbb{E}(C_s(v_1,v_2)) = \Gamma = \phi(v_1, v_2) + \frac{1 - \phi(v_1, v_2)}{m}, \\
    &\mathbb{V}(C_s(v_1,v_2)) = \frac{1 - \Gamma^2}{d} + \frac{2(1-\Gamma)(d-1)}{dm^2}, \\
    &\mathrm{Pr} \left( |C_s(v_1,v_2) - \phi(v_1, v_2) | \geq \eta \Gamma + \frac{1-\phi(v_1, v_2)}{m} \right)\\
    & \quad \quad \quad \quad \leq \frac{1-\Gamma^2}{d \eta^2 \Gamma^2} \textrm{ for any } \eta > 0.
\end{align*}

\textbf{Proof sketch}: Proof consists of analyzing the random variable for the cosine similarity $C_s(E[v_1,:], E[v_2,:])$ and applying Chebyshev's concentration inequality to obtain the tail bounds.\newline

Consider a memory $\mathcal{M}$ of size $m$ initialized randomly by a independent draws of a Bernoulli random variable from $\{-1,+1\}$ with probability 0.5. Then let $a$ be denote the value at any aribitraty memory location in $\mathcal{M}$
Note, for each $k \in \mathbb{N}$
\begin{align}
    E(a^{2k}) = 1 \quad E(a^{2k+1}) = 0
\end{align}
The norm of any embedding of dimension d drawn from this $\mathcal{M}$ is $\sqrt{d}$. Let us look at the inner product

\begin{equation}
    \widehat{\ip{E_x}{E_y}} = \Sigma_{i=1}^d \{ \mathcal{I} (l_i(x) == l_i(y)) \mathcal{M}[l_i(x)]^2 +  \mathcal{I}(l_i(x) != l_i(y)) \mathcal{M}[l_i(x)]\mathcal{M}[l_i(y)] \}
\end{equation}
Let $\Gamma = \phi(x,y) + \frac{1 - \phi(x,y)}{m}$
\begin{equation}
    E(\widehat{\ip{E_x}{E_y}}) = d \Gamma
\end{equation}
\begin{equation}
    E(\widehat{Cosine(E_x, E_y)}) = \Gamma 
\end{equation}
Analysis of Variance is a bit involved due to interdependence between each of the terms in the summation above that comes about due to random collisions caused by rehashing. We analyse both the cases. The case of interdependence just precipitates to independent case for reasonable M.

\textbf{Case 1: Assume independence}
\begin{equation}
    \mathbb{V}(\widehat{\ip{E_x}{E_y}})  =  d (\mathbb{V}( \mathcal{I} (l_i(x) == l_i(y)) +  \mathcal{I}(l_i(x) != l_i(y)) \mathcal{M}[l_i(x)]\mathcal{M}[l_i(y)]))
\end{equation}

\begin{equation}
    \mathbb{V}(\widehat{\ip{E_x}{E_y}})  =  d (\mathbb{V}( \mathcal{I} (l_i(x) == l_i(y)))+
    \mathbb{V}(\mathcal{I}(l_i(x) != l_i(y)) \mathcal{M}[l_i(x)]\mathcal{M}[l_i(y)])))
\end{equation}
\begin{equation}
    \mathbb{V}(\widehat{\ip{E_x}{E_y}})  =  d ( \Gamma(1 - \Gamma) + (1-\Gamma))
\end{equation}
\begin{equation}
    \mathbb{V}(\widehat{\ip{E_x}{E_y}})  =  d  (1+\Gamma)(1 - \Gamma) 
\end{equation}
\begin{equation}
    \mathbb{V}(Cosine(\ip{E_x}{E_y}))  =  \frac{1}{d} (1+\Gamma)(1 - \Gamma) 
\end{equation}
\textbf{Case 2: Don't assume independence.}

\begin{equation}
    \widehat{Cosine(E_x, E_y)} = \frac{1}{d} \widehat{\ip{E_x}{E_y}} 
\end{equation}

\begin{equation}
    \widehat{cosine(E_x, E_y)} = \frac{1}{d} \Sigma_{i=1}^d \{ \mathcal{I} (l_i(x) == l_i(y)) \mathcal{M}[l_i(x)]^2 +  \mathcal{I}(l_i(x) != l_i(y)) \mathcal{M}[l_i(x)]\mathcal{M}[l_i(y)] \}
\end{equation}

\begin{equation}
E(\widehat{Cosine(E_x, E_y)} = \Gamma
\end{equation}

\begin{equation}
E(\widehat{(Cosine(E_x,E_y)}^2) = \frac{1}{d^2} E(\widehat{(\ip{E_x}{E_y})}^2) 
\end{equation}

Let $ex_i = \mathcal{M}[l_i(x)]$
\begin{equation}
E(\widehat{(Cosine(E_x,E_y)}^2)  = \frac{1}{d^2} E (\Sigma ex_i ey_i) (\Sigma ex_i ey_i)
\end{equation}

\begin{equation}
E(\widehat{(Cosine(E_x,E_y)}^2)  = \frac{1}{d^2} E (\Sigma_{i=1}^d (ex_i ey_i)^2  + \Sigma_{i\neq j} (ex_i ey_i) (ex_j ey_j)
\end{equation}

\begin{equation}
E(\widehat{(Cosine(E_x,E_y)}^2)  = \frac{1}{d} ( E ((ex_i ey_i)^2)  + (d-1) E (ex_i ey_i) (ex_j ey_j)
\end{equation}

\begin{equation}
E(\widehat{(Cosine(ex,ey)}^2)  = \frac{1}{d} ( 1  + (d-1) \mathbb{E} \mathcal{M}[l_i(x)]\mathcal{M}[l_j(x)]\mathcal{M}[l_i(y)]\mathcal{M}[l_j(x)])
\end{equation}

\begin{table}[h]
\centering
\begin{tabular}{|l|l|l|l|r|r|}
\hline
 & $l_i(x)$ & $l_i(y)$ & $l_j(x)$ & \multicolumn{1}{l|}{$l_j(y)$} & \multicolumn{1}{l|}{probability} \\ \hline
 & a & a & b & b & pc\textasciicircum{}2           \\ \hline
 & a & b & a & b & (1-pc) / m\textasciicircum{}2   \\ \hline
 & a & b & b & a & (1 - pc) / m\textasciicircum{}2 \\ \hline
\end{tabular}
\caption{Table for reference on cases}
\label{tab:posval}
\end{table}

Using table \ref{tab:posval}, 
\begin{equation}
E(\widehat{(Cosine(ex,ey)}^2)  = \frac{1}{d} ( 1  + (d-1) (\Gamma^2 + 2\frac{( 1 - \Gamma)}{m^2}))
\end{equation}

\begin{equation}
    Var (\widehat{(Cosine(E_x,E_y)}) = E(\widehat{(Cosine(E_x,E_y)}^2 ) - E(\widehat{(Cosine(E_x,E_y)})^2
 \end{equation}

\begin{equation}
    Var (\widehat{(Cosine(ex,ey)}) =\frac{1}{d} ( 1  + (d-1) (\Gamma^2 + 2\frac{( 1 - \Gamma)}{m^2})) - \Gamma^2
 \end{equation}
 
 Collecting terms
 \begin{equation}
    Var (\widehat{(Cosine(ex,ey)}) =\frac{1}{d} \{ (1 - 2\frac{1-\Gamma}{m^2} - \Gamma^2) + 2d(\frac{1-\Gamma}{m^2}) \}
 \end{equation}
 
 \begin{equation}
    Var (\widehat{(Cosine(ex,ey)}) = \frac{1-\Gamma^2}{d} + \frac{2(1-\Gamma)(d-1)}{dm^2} \approx\frac{1}{d}( 1-\Gamma^2)
 \end{equation}
 
 \section{Procedures requires small data sample}
 Assume that the real dataset is of size N, the sparsity of each feature value is s. i.e. the probability of a feature appearing in an example is s. For simplicity, let us assume that each feature value has the same sparsity. This may not be generally true. But nonetheless it helps us draw an idea of data sample needed.  Let us consider the situation where the two features have jaccard similarity J. The venn diagram below shows the distribution of samples in our case. 
 
\begin{figure}[h]
    \centering
    \includegraphics[scale=0.4]{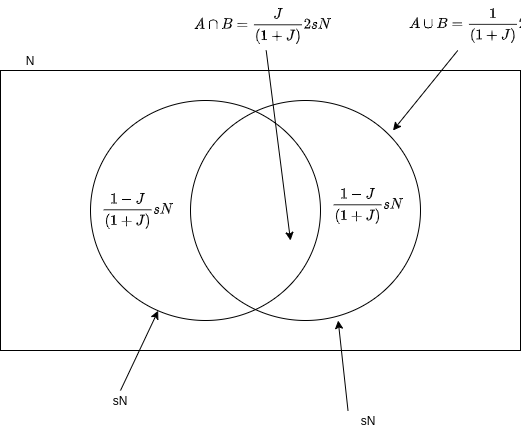}
    \caption{Venn diagram for two features f1,f2}
    \label{fig:my_label}
\end{figure}
Let the events be as follows:
\begin{itemize}
    \item $A_i$ : $i^{th}$ sample has feature f1
    \item $B_i$ : $i^{th}$ sample has feature f2
\end{itemize}
The estimated jaccard similarity after drawing n random i.i.d samples would be,
\begin{equation}
    \hat{J} = \frac{\Sigma_{i=1}^n \mathcal{I}(A_i \wedge B_i)}{\Sigma_{i=1}^n \mathcal{I}(A_i \wedge B_i} = \frac{X}{Y}
\end{equation}
Let X and Y and C be the following
\begin{equation}
    C = \frac{(1+J)}{2ns}
\end{equation}
\begin{equation}
    X = C \Sigma_{i=1}^n \mathcal{I}(A_i \cap B_i) 
\end{equation}
\begin{equation}
    Y = C \Sigma_{i=1}^n \mathcal{I}(A_i \cup B_i) 
\end{equation}
\begin{align}
    E(X) = C n \frac{2sJ}{1+J}  \implies E(X) = \frac{(1+J)}{2ns} n \frac{2sJ}{1+J} = J \\
    E(Y) = C n \frac{2s}{1+J}  \implies E(Y) = \frac{(1+J)}{2ns} n \frac{2s}{1+J} = 1
\end{align}

\begin{align}
    Var(X) = C^2  n \frac{2sJ}{1+J}  (1 - \frac{2sJ}{1+J}) \\
    Var(Y) =C^2  n \frac{2s}{1+J}  (1 - \frac{2s}{1+J}) 
\end{align}
Let us look at Variance of X
\begin{align}
    Var(X) = \frac{(1+J)^2}{4n^2s^2}   n \frac{2sJ}{1+J}  (1 - \frac{2sJ}{1+J}) \\
    Var(X) = \frac{J}{2ns} (1+J -2sJ)
\end{align}

Let us look at Variance of (Y)
\begin{align}
    Var(Y) = \frac{(1+J)^2}{4n^2s^2} n \frac{2s}{1+J}  (1 - \frac{2s}{1+J}) \\
    Var(Y) = \frac{1}{2ns} (1+J -2s) \\
\end{align}

Using Chebysev's inequality , 

\begin{equation}
    \mathbb{P}(|Y - 1| \geq \epsilon) < \frac{Var(Y)}{\epsilon^2}
\end{equation}
    Hence, $1 - \epsilon \leq Y \leq 1 + \epsilon $ with probability $1 - \delta $ where $\delta = \frac{1+J - 2s}{2ns\epsilon^2} $

Hence we can write, with probability $1-\delta$
\begin{equation}
    \frac{X}{1+\epsilon} \leq  \hat{J} \leq \frac{X}{1-\epsilon}
\end{equation}

Hence, with probability $1-\delta$
\begin{equation}
    \frac{J}{1+\epsilon} \leq  E(\hat{J}) \leq \frac{J}{1-\epsilon}
\end{equation}

\begin{equation}
    \frac{X^2}{(1+\epsilon)^2} \leq  \hat{J}^2 \leq \frac{X^2}{(1-\epsilon)^2}
\end{equation}
 \begin{equation}
    \frac{E(X^2)}{(1+\epsilon)^2} \leq  E(\hat{J}^2) \leq \frac{E(X)^2}{(1-\epsilon)^2}
\end{equation}
\begin{equation}
    \frac{Var(X) + E(X)^2}{(1+\epsilon)^2} \leq  E(\hat{J}^2) \leq \frac{Var(X) + E(X)^2}{(1-\epsilon)^2}
\end{equation}
   
\begin{equation}
    \frac{J^2}{(1+\epsilon)^2} \leq  E(\hat{J})^2 \leq \frac{J^2}{(1-\epsilon)^2}
\end{equation}

\begin{equation}
    \frac{Var(X) + E(X)^2}{(1+\epsilon)^2} - \frac{J^2}{(1-\epsilon)^2}  \leq E(\hat{J}^2) - E(\hat{J})^2 \leq \frac{Var(X) + E(X)^2}{(1-\epsilon)^2} - \frac{J^2}{(1+\epsilon)^2}
\end{equation}
\begin{align}
    Var(X) + E(X)^2 = \frac{J}{2ns} (1+J -2sJ + 2nsJ)
\end{align}
The above is the actual result. However for simplicity we simplify assuming $\epsilon$ is small. 

Let 
\begin{equation}
A = Var(X) + E(X)^2 - J^2 
\end{equation}
\begin{equation}
B = Var(X) + E(X)^2 + J^2 
\end{equation}

\begin{equation}
A =\frac{J}{2ns} (1+J -2sJ + 2nsJ) - J^2 
\end{equation}
\begin{equation}
B = \frac{J}{2ns} (1+J -2sJ + 2nsJ) + J^2 
\end{equation}

\begin{equation}
A =\frac{J}{2ns} (1+J -2sJ)
\end{equation}
\begin{equation}
B = A + 2J^2
\end{equation}

\begin{equation}
    A - 2\epsilon (A+ 2J^2) \leq Var(\hat{J}) \leq A + 2\epsilon (A + 2J^2)
\end{equation}
\begin{equation}
    (1-2\epsilon)A - 4\epsilon J^2) \leq Var(\hat{J}) \leq (1+ 2\epsilon) A + 4\epsilon J^2)
\end{equation}
Also under small $\epsilon$

\begin{equation}
    J(1-\epsilon) \leq  E(\hat{J}) \leq J(1+\epsilon)
\end{equation}

\end{document}